\documentclass[conference]{IEEEtran}
\IEEEoverridecommandlockouts
\usepackage{multicol}
\usepackage{lipsum,graphicx,subcaption}
\usepackage{float}
\usepackage{epstopdf}
\usepackage{ifthen}
\usepackage[latin1]{inputenc}

\usepackage{amssymb}
\usepackage[cmex10]{amsmath}
\usepackage{dcolumn}
\usepackage{fixltx2e}
\usepackage[nolist]{acronym}
\DeclareGraphicsExtensions{.pdf,.jpeg,.png,.eps}
\usepackage{booktabs}

\usepackage{comment}
\usepackage{color}

\usepackage{setspace}
\usepackage{mathrsfs}          
\usepackage{amsmath, amssymb, amsthm}
\usepackage{amsfonts}
\usepackage{mathrsfs}
\usepackage{multirow}        
\usepackage{framed}
\usepackage{psfrag}
\usepackage{graphicx}
\usepackage{units}            
\usepackage{algpseudocode} 

\newcolumntype{d}[1]{D{.}{\cdot}{#1} }

\usepackage{bm}
\newcommand{\LeftP} {\left\lbrace }
\newcommand{\RightP}{\right\rbrace }

\newcommand{\LeftPD} {\left( }
\newcommand{\RightPD}{\right)}

\newcommand{\compl}{\mathbb{C}}         

\newcommand{\ma}  [1]{ \bm{#1} } 
\newcommand{\mav}  [1]{ \bm{#1} } 

\newcommand{\NORM}[1]  { \| #1 \|  }
\newcommand{\Ex}[2][]{\mathbb{E}_{#1}\left[ #2\right]} 
\newcommand{\IndexM}[3]{\left[ #1\right]_{\LeftPD #2,#3 \RightPD } } 
\newcommand{\IndexV}[2]{\left[ #1\right]_{\LeftPD #2 \RightPD } } 

\newcommand{\Vect}  [1] {\mathrm{vec}  \LeftP #1 \RightP } 
\newcommand{\diag} [1] {\mathrm{diag} \LeftP #1 \RightP }
\newcommand{\unvec} [3] {\mathrm{unvec}_{#2 \times #3} \LeftP #1 \RightP }
\newcommand{\dft} [1] {\tilde{#1} } 

\newcommand{\DFT} [1] {\ma{F}_{#1}} 
\newcommand{\Tsub} {T_{\text{sub}}} 

\newcommand{\PI}  [2]{ \ma{P}_{#1,#2} }
\newcommand{\V}  [3]{ \ma{V}^{(#3)}_{#1,#2} }
\newcommand{\Z}  [3]{ \ma{Z}^{(#3)}_{#1,#2} }

\newcommand{\Zbar}  [3]{ \bar{\ma{Z}}^{(#3)}_{#1,#2} }

\usepackage{mathtools}
\usepackage{cuted}

\def\BibTeX{{\rm B\kern-.05em{\sc i\kern-.025em b}\kern-.08em
		T\kern-.1667em\lower.7ex\hbox{E}\kern-.125emX}}
\begin{document}
\begin{acronym}
	\acro{OOB}{out-of-band}
	\acro{RRC}{root-raised Cosine}
	\acro{RC}{raised-cosine}
	\acro{ISI}{inter-symbol-interference}
	\acro{ZF}{zero-forcing}
	\acro{MF}{matched filter}
	\acro{SINR}{signal-to-interference-plus-noise ratio}
	\acro{SNR}{signal-to-noise ratio}
	\acro{FIR}{finite impulse repose }
	\acro{DFT}{discrete Fourier transform}
	\acro{OFDM}{orthogonal frequency division multiplexing}
	\acro{GFDM}{generalized frequency division multiplexing}
	\acro{ICI}{inter-carrier-interference}
	\acro{IAI}{inter-antenna-interference}
	\acro{NEF}{noise-enhancement factor}
	\acro{FDE}{frequency fomain equalization}
	\acro{SVD}{singular-value decomposition}
	\acro{AWGN}{additive white Gaussian noise}
	\acro{DTFT}{discrete-time Fourier transform}
	\acro{OFDM}{orthogonal frequency division multiplexing}
	\acro{FFT}{fast Fourier transform}
	\acro{SIR}{signal-to-interference ratio}
	\acro{DZT}{discrete Zak transform}
	\acro{MIMO}{multiple-input multiple-output}
	\acro{PAPR}{peak-to-average power ratio}
	\acro{F-OFDM}{filtered OFDM}
	\acro{CP}{cyclic prefix}
	\acro{CS}{cyclic suffix}
	\acro{ZP}{zero padding}
	\acro{IBI}{inter-block-interference}
	\acro{GT}{guard tone}
	\acro{UF-OFDM}{universal-filtered OFDM}
	\acro{FBMC}{filter bank multicarrier}
	\acro{OQAM}{offset quadrature amplitude modulation}
	\acro{FER}{frame error rate}
	\acro{MMSE}{minimum mean square error}
	\acro{IAI}{inter-antenna-interference}
	\acro{MCS}{modulation coding scheme}
	\acro{PSD}{power spectral density}
	\acro{IoT}{Internet of Things}
	\acro{MTC}{machine-type communication}
	\acro{STC}{space-time coding}
	\acro{TR-STC}{time-reversal space-time coding}
	\acro{MRC}{maximum-ratio combiner}
	\acro{LS}{least squares}
	\acro{LMMSE}{linear minimum mean squared error}
	\acro{CIR}{channel impulse response}
	\acro{STO}{symbol time offset}
	\acro{CFO}{carrier frequency offset}
	\acro{UE}{user equipment}
	\acro{FO}{frequency offset}
	\acro{TO}{time offset}
	\acro{BS}{base station}
	\acro{FMT}{filtered multitone }
	\acro{DAC}{digital-to-analogue converter }
	\acro{FO}{frequency offset}
	\acro{TO}{time offset}
	\acro{ISI}{inter-symbol-interference}
	\acro{IUI}{inter-user-interference}
	\acro{IBI}{inter-block-interference}
	\acro{i.i.d.}{independent and identically distributed}
	\acro{SER}{symbol error rate}
	\acro{LTE}{Long Term Evolution}
	\acro{SISO}{single-input single-output}
	\acro{Rx}{receive}
	\acro{Tx}{transmit}
	\acro{MSE}{mean squared error}
	\acro{IFPI}{interference-free pilot insertion}
	\acro{PDP}{power-delay-profile}
	\acro{ML}{maximum likelihood}
	\acro{5G}{5th generation}
	\acro{4G}{4th generation}
	\acro{NR}{New Radio}
	\acro{eMBB}{enhanced media broadband}
	\acro{URLLC}{ultra-reliable and low-latency communication}
	\acro{mMTC}{massive machine type communication}
	\acro{SDR}{software defined radio}
	\acro{RF}{radio frequency}
	\acro{PHY}{physical layer}
	\acro{MAC}{medium access layer}
	\acro{FPGA}{field programmable gate array}
	\acro{IDFT}{inverse discrete Fourier transform}
	\acro{DRAM}{dynamic random access memory}
	\acro{BRAM}{block RAM}
	\acro{FIFO}{first in first out}
	\acro{D/A}{digital to analog}
\acro{EVA} {extended vehicular A channel model} 
	\acro{OTFS}{Orthogonal time frequency space modulation}
	\acro{SFFT}{symplectic finite Fourier transform}

	\acro{ACLR}{adjacent channel leakage rejection}
	\acro{ADC}{analog-to-digital converter}
	\acro{AGC}{automatic gain control}
	\acro{CEP}{channel estimation preamble}
	\acro{DPD}{digital pre-distortion}
	\acro{PA}{power amplifier}
		\acro{LTV}{linear time-variant}
	
		\acro{NMSE}{normalized mean-squared error}
	\acro{PRB}{physical resource block}
	\acro{BER}{bit error rate}
		\acro{FER}{frame error rate}
	\acro{DL}{downlink}
	\acro{UL}{uplink}
	\acro{FO}{frequency offset}
	\acro{TO}{time offset}
	\acro{MA}{multiple access}

	\acro{INI}{inter-numerology-interference}
	\acro{PCCC}{parallel concatenated convolutional code}
	\acro{CCDF}{complementary cumulative distribution function}
	\acro{SC}{single carrier}
	\acro{FDMA}{frequency division multiple access}
\end{acronym}
\title{Extended GFDM Framework: OTFS and GFDM Comparison\\	
	\thanks{The work presented in this paper has been performed in the framework of the ORCA project [https://www.orca-project.eu/].This project has received funding from the Eropean Union's Horizon 2020 research and innovation programme under grant agreement No 732174.}}
\author{
\IEEEauthorblockN{Ahmad Nimr, Marwa Chafii, Maximilian Matth\'{e},    Gerhard Fettweis}
\IEEEauthorblockA{ Vodafone Chair Mobile Communication Systems, Technische Universit\"{a}t Dresden, Germany}
\IEEEauthorblockA{\small\texttt{\{first name.last name\}@ifn.et.tu-dresden.de}}
}

\maketitle
\IEEEpeerreviewmaketitle
\begin{abstract}
\ac{OTFS} has been recently proposed to achieve  time and frequency diversity, especially in \ac{LTV} channels with large Doppler frequencies. The idea is based on the precoding of the data symbols using \ac{SFFT} then transmitting them by mean of \ac{OFDM} waveform. Consequently, the demodulator and channel equalization can be coupled in one processing step. As a distinguished feature, the demodulated data symbols have roughly equal gain independent of the channel selectivity. On the other hand,  \ac{GFDM} modulation also employs the spreading over the time and frequency domains using circular filtering. Accordingly, the data symbols are implicitly precoded in a similar way as applying \ac{SFFT} in \ac{OTFS}. In this paper, we present an extended representation of \ac{GFDM} which shows that  \ac{OTFS} can be processed as a \ac{GFDM} signal with simple permutation. Nevertheless, this permutation is the key factor behind the outstanding performance of OTFS in \ac{LTV} channels, as demonstrated in this work.  Furthermore, the representation of \ac{OTFS} in the \ac{GFDM} framework provides an efficient implementation, that has been intensively investigated for \ac{GFDM}, and facilitates the understanding of the \ac{OTFS} distinct features.
\end{abstract}

\acresetall
\section{Introduction}\label{sec:introduction}
In the contention between 5G waveform candidates, several modulation techniques were proposed to attain the requirements for different use cases.  Some of the waveforms focus on providing very low \ac{OOB} emission, e.g. \ac{FBMC} \cite{FBMC}, which allows asynchronous multiple access. Other designs  concern about the implementation complexity, which leads to several proposals based on \ac{OFDM}, such as windowed-\ac{OFDM} \cite{medjahdi2017wola}, filtered-\ac{OFDM} \cite{wild20145g}. Furthermore, \ac{GFDM} \cite{GFDM} was first proposed as an alternative to \ac{OFDM} to improve the spectral efficiency by reducing the \ac{CP} overhead of \ac{OFDM}. However, due to the flexibility in tuning the different parameters, such as the number of subcarriers and subsymbols, the prototype pulse shape and the active subsymbol set, \ac{GFDM} can be reconfigured to meet different requirements \cite{zhang2017study}. This flexibility inspires the design and implementation of unified multicarrier framework based on the \ac{GFDM} model.

\ac{OTFS} modulation technique has been recently proposed in \cite{OTFS_wcnchadani2017orthogonal} to deal with high mobility scenarios. In this approach, the data symbols are spread in the time and frequency domains using \ac{SFFT}-based precoding. As a consequence, high diversity gain is achieved. The spread data symbols are then transmitted with \ac{OFDM} waveform. Compared to \ac{OFDM}, a very significant gain in terms of \ac{FER} is shown in high mobility case \cite{OTFS_wcnchadani2017orthogonal}. Furthermore, the processing is of relatively  low cost, where the channel equalization and demodulation are coupled at the receiver. Interestingly, all the estimated symbols have the same \ac{SNR} at the output of the equalizer. This feature can not be achieved with  \ac{OFDM} unless power allocation techniques are applied. However, this requires the channel knowledge at the transmitter, which is infeasible under higher Doppler frequencies.

Because \ac{GFDM} is based on circular filtering in the time and frequency domains, the data symbols are implicitly spread in a similar way as in \ac{OTFS}. This motivates us to investigate  the structure of \ac{GFDM} and reveal a relation between both systems. As we show in this paper, the \ac{OTFS} samples can be generated from a permutation of \ac{GFDM} signal, which exchanges the role of the number of  subcarriers in \ac{OTFS} to be the number of subsymbols in the \ac{GFDM} terminology and vice versa. This observation is relevant for two reasons; first, a slight modification of a real-time  flexible implementation of \ac{GFDM} transceiver \cite{danneberg2015flexible},  enables the processing of \ac{OTFS}. Second, the properties of \ac{OTFS} can be clearly derived from the structure of \ac{GFDM}. For example, we show that the coupled equalizer and demodulator are equivalent to circular filtering of the received \ac{OTFS} signal using a \ac{GFDM} receive filter computed from the transmit pulse and the estimated channel.

The remainder of the paper is organized as follows: in Section \ref{sec:GFDM review} we provide a short overview of the conventional \ac{GFDM}. Section \ref{sec: advanced matrix model} is dedicated to the advanced representation of \ac{GFDM} and extended framework. The relation between the conventional \ac{GFDM} and \ac{OTFS} is discussed in Section \ref{sec:OTFS in GFDM}. Numerical comparisons through simulation are introduced in Section \ref{sec:numerical example }. Finally, Section \ref{sec:conclusions} concludes the paper.
\section{GFDM overview}\label{sec:GFDM review}
In the \ac{GFDM} modulation, a time-frequency resource block of duration $T$ and bandwidth $B$ is used to convey a message of maximum $N$ data symbols. For that, the frequency band is divided into $K$ equally spaced subcarriers with subcarrier spacing $\Delta f = \tfrac{B}{K}$, and the time duration  is divided into $M$ equally spaced subsymbols with subsymbol spacing  ${T_{\text{sub}} = \tfrac{T}{M}}$ such that $\Delta f T_{\text{sub}} = 1$.
Each pair of subcarrier-subsymbol $(k,m)$ is used to modulate one data symbol $d_{k,m}$ using a pulse shape $g_{k,m}(t)$, and thus,  $N = MK $. The pulse shapes are generated by a shift in the time and frequency domains of a  periodic prototype pulse shape $g_T(t)$, in addition to windowing to confine the block within the duration $T$, thus
\begin{equation}
\small
g_{k,m}(t) = w_T(t)g_{T}(t-m\Tsub)e^{j2\pi k\Delta f t}, \label{eq: pulse shapes no cp}
\end{equation}
where $w_T(t)$ is a rectangular window of duration $T$. Accordingly, one \ac{GFDM} block in the time domain is expressed as 
\begin{equation}
\small
x(t) = w_T(t)\sum\limits_{k=-K/2}^{K/2-1}\sum\limits_{m = 0}^{M-1}\ d_{k,m} g_{T}(t-m\Tsub)e^{j2\pi k\Delta f t}. \label{eq:GFDM continious}
\end{equation}
based on this representation, \ac{GFDM} can be seen as generalization of \ac{OFDM}, where $g_T(t) = 1$ and $M=1$. 
The discrete-time representation can be achieved by using a sampling frequency $F_s = B$. The samples of one \ac{GFDM} block can be represented in a vector  $\ma{x} \in \compl^{N\times 1}$ such that \cite{GFDM},
\begin{equation}
\small
\IndexV{\ma{x}}{n} = x(\tfrac{n}{B}) = \sum\limits_{k=0}^{K-1}\sum\limits_{m = 0}^{M-1} d_{k,m} g[<n -mK>_N] e^{j2\pi\frac{k}{K}n}. \label{eq: Discrete model}
\end{equation}
Here, $<.>_N$ denotes the modulo-$N$ operator. The corresponding representation in the frequency domain is given by
\begin{equation}
\small
\IndexV{\tilde{\ma{x}}}{n}= \sum\limits_{k=0}^{K-1}\sum\limits_{m = 0}^{M-1} d_{k,m} \tilde{g}[<n -kM>_N] e^{-j2\pi\frac{m}{M}n}. \label{eq: Discrete model frequency}
\end{equation}
The \ac{GFDM} demodulator applies circular filtering on  the received signal $y[n]$ with a receiver pulse $\ma{\gamma}[n]$ such that, 
\begin{equation}
\small
\hat{d}_{k,m} = \sum_{n=0}^{N-1}y[n]\gamma^*[<n -mK>_N] e^{-j2\pi\frac{k}{K}n}.\label{eq:demodulator discrete}
\end{equation}
In the conventional matrix representation, the \ac{GFDM} block can be expressed using a matrix $\ma{A}\in \compl^{N\times N}$ as 
\begin{equation}
\small
\begin{split}
\ma{x} &= \ma{A}\ma{d},
\IndexM{\ma{A}}{n}{k+mK} = g[<n-mK>_N]e^{j2\pi\frac{k}{K}n},
\end{split}
\end{equation}
where $\ma{d}= \Vect{\ma{D}}$ with $\IndexM{\ma{D}}{k}{m} = d_{k,m}$. Moreover, the demodulation matrix $\ma{B}\in \compl^{N\times N}$ has a similar structure and is generated from the receiver pulse $\gamma$  as 
\begin{equation}
\small
\IndexM{\ma{B}}{n}{k+mK} = \gamma[<n-mK>_N]e^{j2\pi\frac{k}{K}n},~ \hat{\ma{d}} = \ma{B}^H\ma{y}.
\end{equation}
\section{GFDM alternative representation }\label{sec: advanced matrix model}
\begin{figure}[h]
	\centering
	\includegraphics[width=.58\linewidth]{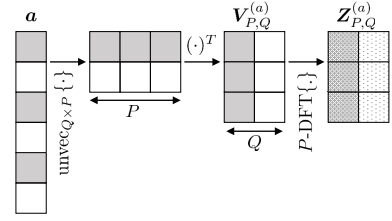}
	\caption{Visualization of \eqref{eq:reshape x} and \eqref{eq:ZAK}.}\label{fig:visualization of zak}
\end{figure}

In this section, we introduce an alternative matrix model by arranging the \ac{GFDM} block $\ma{x}$ in a matrix instead of a vector. Initially, we define the following  auxiliary matrices for a generic vector $\ma{a}\in \compl^{PQ\times 1}$ as visualized in Fig.~\ref{fig:visualization of zak},
\begin{align}
\small
\V{P}{Q}{\mav{a}}&= \unvec{\mav{a}}{Q}{P} ^T\Leftrightarrow\IndexM{\V{P}{Q}{\mav{a}}}{p}{q} = \IndexV{\ma{a}}{q+pQ}\label{eq:reshape x}, \\
\Z{P}{Q}{\mav{a}}&= \DFT{P}\V{P}{Q}{\mav{a}} , ~
\Zbar{Q}{P}{\tilde{\mav{a}}}= \frac{1}{Q}\DFT{Q}^H\V{Q}{P}{\tilde{\mav{a}}} , \label{eq:ZAK}
\end{align}
where $\tilde{\ma{a}} = \DFT{PQ}\ma{a} $, $\unvec{\mav{a}}{Q}{P}$ denotes the  inverse of vectorization operation, $\DFT{P}$  is the $P$-point \ac{DFT} matrix, defined by $\IndexM{\DFT{P}}{i}{j} = e^{-j2\pi\frac{ij}{P}}$.
The matrix $\Z{P}{Q}{\mav{a}}$,  $\Zbar{Q}{P}{\tilde{\mav{a}}}$ are known as the \ac{DZT} \cite{Bolcskei1997} of $\ma{a}$ and $\tilde{\ma{a}}$, respectively.
\subsection{Time domain representation}
The \ac{GFDM} block equation \eqref{eq: Discrete model} can be reformulated by representing $n$ with two indexes $q = 0,\cdots, K-1$ and $p = 0\cdots, M-1$, such that $n = q + pK$. Thereby, 
\begin{align*}
\small
\IndexV{\ma{x}}{q+pK}
&= \sum_{m=0}^{M-1}\sum_{k=0}^{K-1} d_{k,m} g[<q+pK -mK>_N] e^{j2\pi\frac{k}{K} q}
\end{align*}
Using the notations in \eqref{eq:reshape x}, then 
\begin{align*}
\small
\begin{split}
\IndexM{\V{M}{K}{\ma{x}}}{p}{q}
&= \sum_{m=0}^{M-1}\IndexM{\V{M}{K}{\ma{g}}}{<p-m>_M}{q}\sum_{k=0}^{K-1} \IndexM{\ma{D}}{k}{m}  e^{j2\pi\frac{k}{K} q}\\
&= \sum_{m=0}^{M-1}\IndexM{\V{M}{K}{\ma{g}}}{<p-m>_M}{q}\IndexM{\ma{D}^T\DFT{K}^H}{m}{q}.
\end{split} \label{eq:reformulated x}
\end{align*}
The second line defines a circular convolution between the $q$-th column of $\V{M}{K}{\ma{g}}$ and the $q$-th column of $\ma{D}^T\DFT{K}^H$, which can be expressed in the frequency domain with $M$-\ac{DFT} as
\begin{equation}
\small
\begin{split}
\IndexM{\DFT{M}\V{M}{K}{\ma{x}}}{p}{q} =
\IndexM{\DFT{M}{\ma{V}}_{M,K}^{(\ma{g})}}{q}{q}\cdot \IndexM{\DFT{M}\ma{D}^T\DFT{K}^H}{p}{q}\,\label{eq:x_vect gfdm}.
\end{split}
\end{equation}
Using the notation in \eqref{eq:ZAK}, we  get
\begin{equation}
\small
\begin{split}
{\V{M}{K}{\ma{x}}} =
\frac{1}{M K}\DFT{M}^H\left(K\Z{M}{K}{\ma{g}}\odot\left[\DFT{M}{\ma{D}^T\DFT{K}^H}\right]\right). \label{eq: Time domain model}
\end{split}
\end{equation}
Here, $\odot$ denotes the element-wise multiplication operator.
\subsection{Frequency domain representation}
Following similar derivation on $\dft{\ma{x}}$ defined in \eqref{eq: Discrete model frequency}, we get
\begin{equation*}
\small
\begin{split}
\IndexM{\V{K}{M}{\tilde{\ma{x}}}}{q}{p}
&= \sum_{k=0}^{K-1}\IndexM{\V{K}{M}{\tilde{\ma{g}}}}{<q-k>_K}{p}\sum_{m=0}^{M-1} \IndexM{\ma{D}}{k}{m}   e^{-j2\pi\frac{m}{M} p}. \label{eq:frequency-domain}
\end{split}
\end{equation*}
\begin{equation}
\small
\begin{split}
\mbox{Hence, }{\V{K}{M}{\tilde{\ma{x}}}} =
\frac{1}{K}\DFT{K}\left(K\Zbar{K}{M}{\tilde{\ma{g}}}\odot\left[\DFT{K}^H{\ma{D}\DFT{M}}\right]\right). \label{eq: frequency domain model}
\end{split}
\end{equation}
\subsection{New interpretation of GFDM }
\begin{figure*}[h]
	\centering
	\includegraphics[width=.9\linewidth]{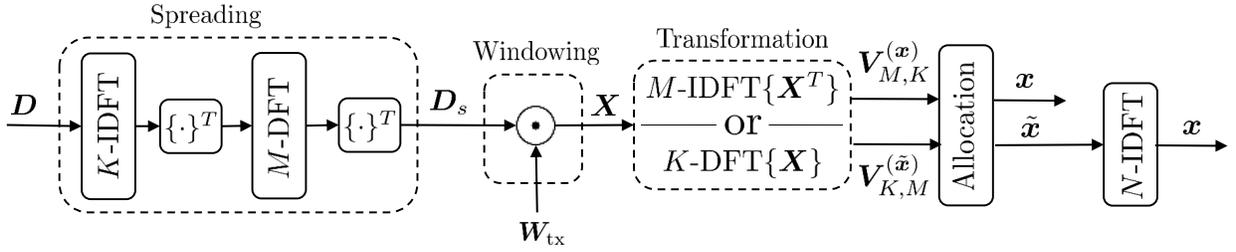}
	\caption{\ac{GFDM} modulator in 4 steps}\label{fig:Modulator }
\end{figure*}
Based on \eqref{eq: Time domain model} and \eqref{eq: frequency domain model}, \ac{GFDM} modulation can be split onto four steps, as shown in Fig.~\ref{fig:Modulator }
\subsubsection{Data spreading }
 the spreading is achieved by applying \ac{DFT} on the rows and \ac{IDFT} on the columns of $\ma{D}$. Therefore, we get the spread data matrix 
\begin{equation}
\ma{D}_s = \frac{1}{K}\DFT{K}^H\ma{D}\DFT{M}. 
\end{equation}
\subsubsection{Windowing}
the spread data matrix $\ma{D}_s$ is element-wise multiplied with a transmitter windowing matrix ${\ma{W}_{\text{tx}} \in \compl^{K\times M}}$, which is generated based on the prototype pulse shape: $\ma{X} = \ma{W}_{\text{tx}} \odot \ma{D}_s$.
The entries of $\ma{W}_{\text{tx}} $ depends on the implementation domain,  
\begin{equation}
\small
\begin{array}{cc}
\ma{W}^{(\text{TD})}_{\text{tx}} = K{\Z{M}{K}{\ma{g}}}^T &,  \ma{W}^{(\text{FD})}_{\text{tx}} = K{\Z{K}{M}{\tilde{\ma{g}}}}.
\end{array}
\end{equation}
\subsubsection{Transformation} \label{sub: transformation}
the matrix $\ma{X}$ can be seen as frequency domain blocks of length $M$ in rows, or as time domain symbols of length $K$  in columns. Thus, a final  $M$-\ac{DFT}  or $K$-\ac{IDFT} is applied to obtain the block samples, as follows
\begin{equation}
\small
\begin{array}{cc}
{\V{M}{K}{\ma{x}}} = \frac{1}{M}\DFT{M}^H\ma{X}^T,& {\V{K}{M}{\tilde{\ma{x}}}} = \DFT{K}\ma{X}
\end{array}.\label{eq:GFDM VX matrix}
\end{equation}
\subsubsection{Allocation}\label{sub: mapping}
the final vector is achieved by allocating the generated samples to the corresponding indexes. Specifically, 
\begin{equation}
\small
\begin{array}{cc}
	\ma{x} = \Vect{{\V{M}{K}{\ma{x}}}^T},& \tilde{\ma{x}} = \Vect{{\V{K}{M}{\tilde{\ma{x}}}}^T}.
\end{array}
\end{equation}
Additionally,  $N$-\ac{IDFT} is required for the frequency domain implementation so the time domain signal is $\ma{x} = \frac{1}{N}\DFT{N}^H\tilde{\ma{x}}$.
The demodulator performs the inverse steps. First, the matrix $\ma{Y}\in \compl^{K\times M}$ is constructed from the received  signal $\ma{y}_{\text{eq}}$,
\begin{equation}
\small
\begin{array}{cc}
\ma{Y}^{(\text{TD})} =  \left[\DFT{M}\V{M}{K}{{\ma{y}_{\text{eq}}}}\right]^T,& \ma{Y}^{{(\text{FD})}} =  \frac{1}{K}\DFT{K}^H\V{K}{M}{\tilde{\ma{y}}_{\text{eq}}}.
\end{array}
\end{equation}
Then a receive window $\ma{W}_{\text{rx}}$ is applied to $\ma{Y}$ followed by despreading, so that $\hat{\ma{D}} = \frac{1}{M }\DFT{K}\left(\ma{W}_{\text{rx}}\odot\ma{Y}\right)\DFT{M}^H$, which
is an alternative representation of the circular demodulation in  \eqref{eq:demodulator discrete}. Table \ref{Tab: 4 steps} summarizes all the processing steps.
\subsection{Extended flexibility of GFDM }
In addition to the main configuration parameters  $K$, $M$ and  the prototype pulse shape, further degrees of freedom can be exploited to extend the flexibility of \ac{GFDM} modem, namely:
\begin{itemize}
\item Flexible spreading: the spreading can be altered to enable or disable the spreading matrices, as in \cite{matthe2016precoded}. 
\item  Flexible transformation: the \ac{DFT} or \ac{IDFT} transformation can be turned on or off to get more options. 
\item Flexible allocation:
the conventional allocation preserves the spectral characteristics of the \ac{GFDM} signal. However, the mapping can be customized by changing the distribution of the samples in the transmitted signal. This extended flexibility is exploited in the next section to generate the \ac{OTFS} signal.
\end{itemize}
\begin{table*}
	\centering
	\small
	\caption{The four steps of conventional \ac{GFDM} modulator.}\label{Tab: 4 steps}
	\begin{tabular}{l|c|c|c|c}	 
		\small	
		\rule{0pt}{12pt}	 & \multicolumn{2}{c|}{Modulation}& \multicolumn{2}{c}{Demodulation}\\
		\cline{2-5}
		\rule{0pt}{12pt}	 & TD implementation&  FD implementation & TD implementation&  FD implementation\\ 
		\hline 
		\rule{0pt}{10pt}	Spread. & \multicolumn{2}{c|}{$\ma{D}_s = \frac{1}{K}\DFT{K}^H\ma{D}\DFT{M}$ } & \multicolumn{2}{c}{$\hat{\ma{D}} = \frac{1}{M}\DFT{K}\hat{\ma{D}}_s\DFT{M}^H$ } \vspace{2pt}\\ 
		\hline 
		\hspace{3pt}\rule{0pt}{13pt}\multirow{2}{*}{Wind.}	 	& $ \ma{W}_{\text{tx}} = K\{\Z{M}{K}{\ma{g}}\}^T$ &  $\ma{W}_{\text{tx}} = K{\Zbar{K}{M}{\tilde{\ma{g}}}}$& \multicolumn{2}{c}{$\ma{W}_{\text{rx}}$ generated from $\ma{W}_{\text{tx}}$} \\ 
		\cline{2-5}
		\rule{0pt}{11pt}	  & \multicolumn{2}{c|}{$\ma{X} = \ma{W}_{\text{tx}} \odot \ma{D}_s$ } & \multicolumn{2}{c}{$\hat{\ma{D}}_s = \ma{W}_{\text{rx}} \odot \ma{Y}$ }\\ 
		\hline 
		\rule{0pt}{16pt}	Trans. & ${\V{M}{K}{\ma{x}}} = \frac{1}{M}\DFT{M}^H\ma{X}^T$& ${\V{K}{M}{\tilde{\ma{x}}}} = \DFT{K}\ma{X}$  & $\ma{Y} =  \left[\DFT{M}\V{M}{K}{{\ma{y}_{\text{eq}}}}\right]^T$& $\ma{Y} =  \frac{1}{K}\DFT{K}^H\V{K}{M}{\tilde{\ma{y}}_{\text{eq}}}$\vspace{2pt}\\
		\hline 
		\rule{0pt}{15pt}	Alloc. & 	$\ma{x} = \Vect{\{\V{M}{K}{\ma{x}}\}^T}$& $\tilde{\ma{x}} = \Vect{\{\V{K}{M}{\tilde{\ma{x}}}\}^T}$&$ \IndexM{\V{M}{K}{\ma{y}_{\text{eq}}}}{m}{k} = \IndexV{\ma{y}_{\text{eq}}}{k+mK}$&$ \IndexM{\V{K}{M}{\tilde{\ma{y}}_{\text{eq}}}}{k}{m} = \IndexV{\tilde{\ma{y}}_{\text{eq}}}{m+kM}$\vspace{0pt}\\
		\hline 		
	\end{tabular}
\end{table*}

\section{OTFS in the GFDM framework} \label{sec:OTFS in GFDM}
\subsection{OTFS overview}
In the \ac{OTFS}  modulation \cite{OTFS_wcnchadani2017orthogonal}, a   packet burst of duration ${T = N_oT_o}$ and bandwidth $B = M_o{\Delta f}_o $  is used to transmit the data symbols  $\ma{D}_o \in \compl^{N_o\times M_o}$. Although \ac{OTFS} is originally designed in the delay-Doppler domain, it can be simply introduced in the time-frequency representation with three-step processing. First, the data symbols are  spread with the inverse of \ac{SFFT} to generate the spread data matrix  $\ma{D}_{s_o} $  defined by \footnote{The subscript $o$ is added to \cite{OTFS_wcnchadani2017orthogonal} to distinguish the OTFS terminology.}
\begin{equation}
	\ma{D}_{s_o} = \frac{1}{N_o} \DFT{N_o}^H \ma{D_o} \DFT{M_o} \in \compl^{N_o\times M_o}.\label{eq:DP}
\end{equation}
Then, a transmit window $\ma{W}_{\text{tx}}\in\compl^{N_o\times M_o}$ can be applied, 
\begin{equation}
	\ma{X}_o = \ma{W}_{\text{tx}}\odot \ma{D}_{s_o}.\label{eq:Xo}
\end{equation}
Finally, the time-domain signal is generated with a multicarrier modulation of $M_o$ subcarriers and $N_o$ subsymbols as
\begin{equation*}
	s(t) = \sum\limits_{m=-\tfrac{M_o}{2}}^{\tfrac{M_o}{2}-1}\sum\limits_{n = 0}^{N_o-1}\IndexM{\ma{X}_o}{n}{m} g_{\text{tx}}(t-nT_o)e^{j2\pi m{\Delta f}_o (t-nT_o)}. \label{eq:OTFS continious}
\end{equation*}
Despite of its general representation, it is implicitly considered, as stated in the works related to \ac{OTFS}, e.g. \cite{OTFS_implementation_farhang2017low}, that 	$g_{\text{tx}}(t)$ is a rectangular pulse of duration $T_o$ and  $T_o{\Delta f}_o = 1$. Therefore, 
the discrete \ac{OTFS} block is given by
\begin{equation*}
	\IndexV{\ma{s}}{l} = \sum\limits_{m=0}^{M_o-1}\sum\limits_{n=0}^{N_o-1}\IndexM{\ma{X}_o^T}{m}{n} g_{\text{tx}}[<l-nM_o>_{(N_oM_o)}]  e^{j2\pi \frac{ml}{M_o}}. \label{eq: OTFS in GFDM}
\end{equation*}
This shows that the \ac{OTFS} block is composed of  $N_o$ consecutive \ac{OFDM} symbols of length $M_o$. Here, the \ac{CP} is not considered. Let ${\V{M_o}{N_o}{\ma{x}_o} = \frac{1}{M_o}\DFT{M_o}^H\ma{X}_o^T}$, with ${\ma{x}_o = \Vect{\{\V{M_o}{N_o}{\ma{x}_o}\}^T}}$, then
\begin{equation}
\ma{s} = \Vect{\V{M_o}{N_o}{\ma{x}_o}} .\label{eq:OTFS signal}
\end{equation} 
\subsection{OTFS relation to GFDM}
Comparing the \ac{OTFS} samples  \eqref{eq:OTFS signal} with the \ac{GFDM} samples \eqref{eq:GFDM VX matrix}, we reveal that the \ac{OTFS} samples can be generated using the first three steps of the  \ac{GFDM} modulator with the parameters $M = M_o$, $K = N_o$.
The difference between both blocks is in the allocation step. Actually, the  \ac{OTFS} block is a permutation of the \ac{GFDM} block $\ma{x}_o$ with  the square commutation matrix  $\PI{N_o}{M_o}\in \Re^{N_oM_o \times N_oM_o}$, i.e. ${\ma{s} = \PI{N_o}{M_o}\ma{x}_o}$, 
\begin{equation}
\small 
\mbox{where } \ma{x}_0 = \Vect{{\V{M_o}{N_o}{\ma{x}_o}}^T} = \PI{N_o}{M_o}^T \Vect{\V{M_o}{N_o}{\ma{x}_o}}.
\end{equation}
\subsection{Received signal model}
The \ac{CP} insertion is performed per block in the case of \ac{GFDM}, while a \ac{CP} is added to each \ac{OFDM} symbol in \ac{OTFS}. The received \ac{OTFS} block $\ma{r}$ after removing the \ac{CP} from each \ac{OFDM} symbol can be expressed in a matrix form as
\begin{equation*}
\small
\begin{split}
\V{M_o}{N_o}{\ma{y}_o} &= \unvec{\ma{r}}{M_o}{N_o}, \\
\ma{y}_0 &= \Vect{{\V{M_o}{N_o}{\ma{y}_o}}^T} = \PI{N_o}{M_o}^T\ma{r}.
\end{split}
\end{equation*}
Consider a \ac{LTV} channel with the response $h(l,n)$ of $L$ delay taps. Following the derivation steps in the appendix on each \ac{OFDM} symbol, we get
\begin{equation}
\small
\begin{split}
\DFT{M_o}\V{M_o}{N_o}{\ma{y}_o}& = \tilde{\ma{H}}_o^{(e)}\odot  \ma{X}_o^T + {\ma{E}_{d_o}} +\ma{V}_o \label{eq:channel OTFS}
\end{split}
\end{equation}
where $\tilde{\ma{H}}^{(e)}_o \in \compl^{M\times K}$ is the equivalent channel defined by
\begin{equation*}
\small
\IndexM{\tilde{\ma{H}}_o^{(e)}}{p}{q} = \frac{1}{M_o}\sum_{m=0}^{M_o-1}\sum_{l=0}^{L-1} h(l,N_{cp}+m+q{M}^{(cp)}_o)e^{-j2\pi\frac{lp}{M_o}},
\end{equation*}
$\small {M}^{(cp)}_o = M_o + N_{cp}$ is the length of the CP-\ac{OFDM} subsymbol, $\ma{E}_{d_o}$ denotes the interference terms arises from the Doppler spread and $\ma{V}_o$ is the additive noise samples.  Furthermore, let 
\begin{equation}
\small
\begin{split}
\ma{W}^{(e)}_{\text{tx}} = \tilde{\ma{H}}_o^{(e) T}\odot \ma{W}_{\text{tx}},~
\ma{X}_o^{(e)} = \ma{W}^{(e)}_{\text{tx}}\odot\ma{D}_{s_o}, \label{eq:eq_window}
\end{split}
\end{equation} 
\begin{equation}
\small\begin{split}
\mbox{then }
\V{M_o}{N_o}{\ma{y}_o} &= \frac{1}{M_o}\DFT{M_o}^H\ma{X}_o^{(e)T}+ \ma{E}_{d_o} +\ma{V}_o\\
&= \V{M_o}{N_o}{\ma{x}^{(e)}_o}+\ma{E}_{d_o} +\ma{V}_o
\end{split}
\end{equation}
where, $\ma{x}^{(e)}_o$ is an equivalent time-domain \ac{GFDM} block generated by the window $\ma{W}^{(e)}_{\text{tx}}$ \eqref{eq:GFDM VX matrix}. Furthermore, the relation
\begin{equation}
\small
\begin{split}
\ma{y}_o = \PI{N_o}{M_o}^T\cdot  \ma{r} =  \ma{x}^{(e)}_o+\ma{\epsilon}_{d_o} +\ma{v}_o,
\end{split}
\end{equation}
shows that the received \ac{OTFS} block can  be processed with a ready implemented real-time \ac{GFDM} receiver, e.g.  \cite{danneberg2015flexible}, under  additive noise channel. In this case, the \ac{MMSE} receive pulse shape can be efficiently calculated based on the inverse Zak transform of $\ma{W}^{(e)}_{\text{tx}}$ \cite{Gabor}.  Moreover, due to the  circular filtering in the combined equalization and demodulation, the symbols at the output achieve roughly the same \ac{SNR} considering the additional interference.
The received \ac{GFDM} block after removing the \ac{CP} is:
\begin{align}
\small
\tilde{\ma{y}} = \tilde{\ma{h}}^{(e)}\odot  \tilde{\ma{x}} + \tilde{\ma{\epsilon}}_d +\tilde{\ma{v}},\label{eq:channel GFDM}
\end{align}
\begin{equation}
\small
\mbox{ where }~ \IndexV{\tilde{\ma{h}}^{(e)}}{q} = \frac{1}{N}\sum_{n=0}^{N-1}\sum_{l=0}^{L-1} h(l,N_{cp}+n)e^{j2\pi\frac{lq}{N}}.
\end{equation} 
Thereby, additional channel equalization processing is required. However, the overall equalization and demodulation is not necessarily equivalent to circular filtering  and the data symbols at the output of the demodulator have different \acp{SNR}.

\begin{figure*}
	\centering
	\begin{subfigure}[b]{0.31\textwidth}
		\includegraphics[width=\textwidth]{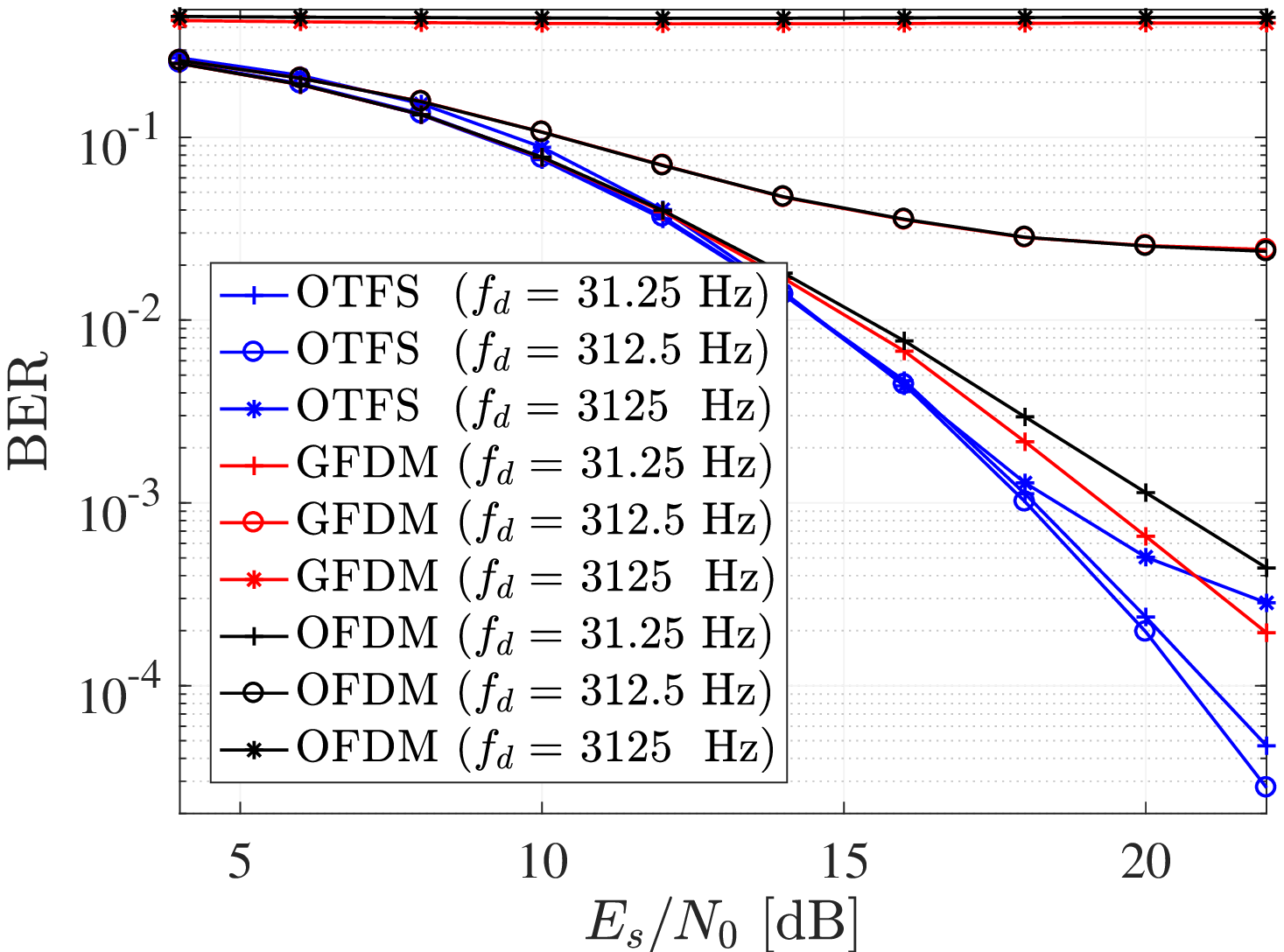}
		\caption{BER vs. SNR.}
		\label{fig:BER_l}
	\end{subfigure}
	~ 
	\begin{subfigure}[b]{0.31\textwidth}
		\includegraphics[width=\textwidth]{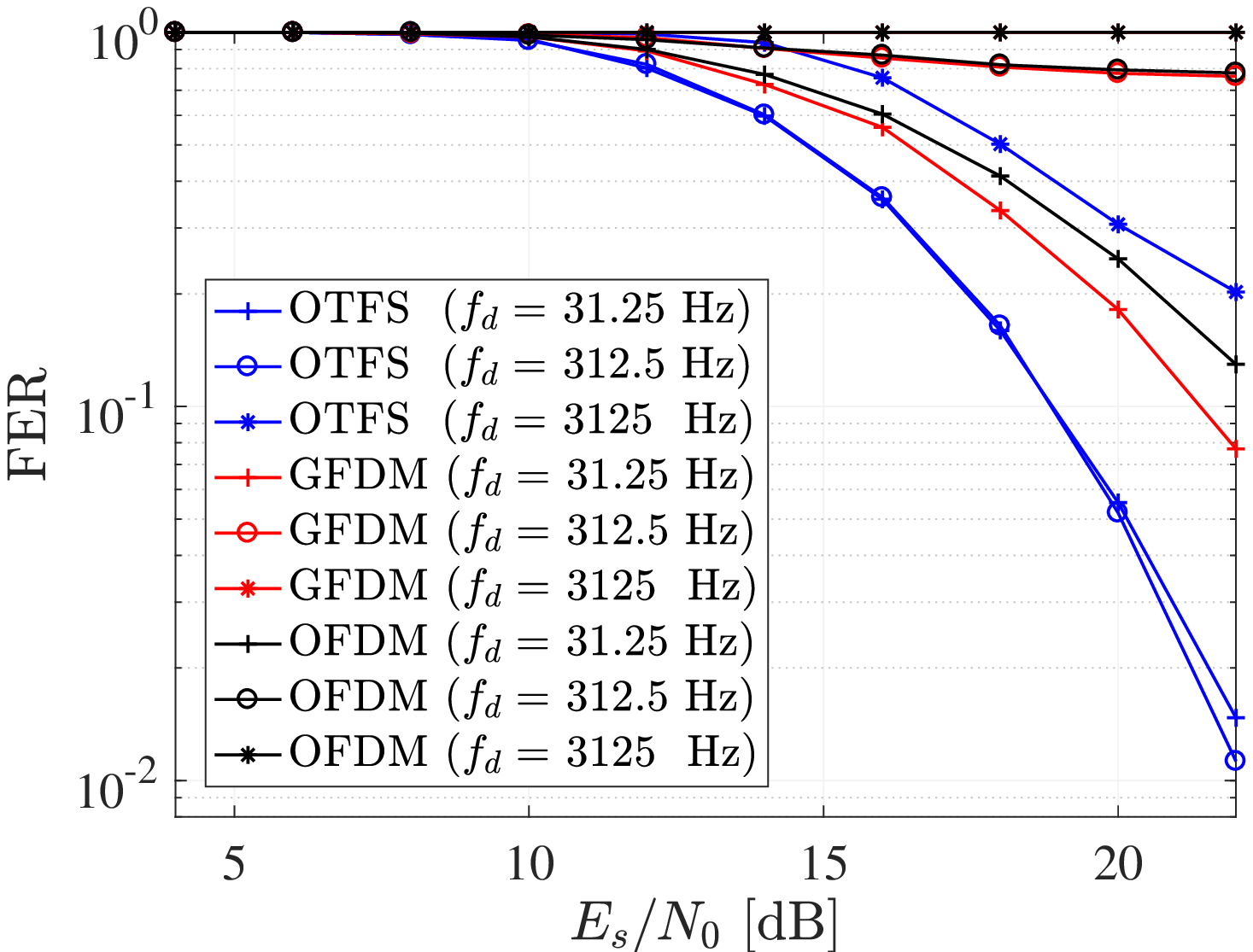}
		\caption{FER vs. SNR.}
		\label{fig:FER_l}
	\end{subfigure}
	~ 
		\begin{subfigure}[b]{0.31\textwidth}
		\includegraphics[width=\textwidth]{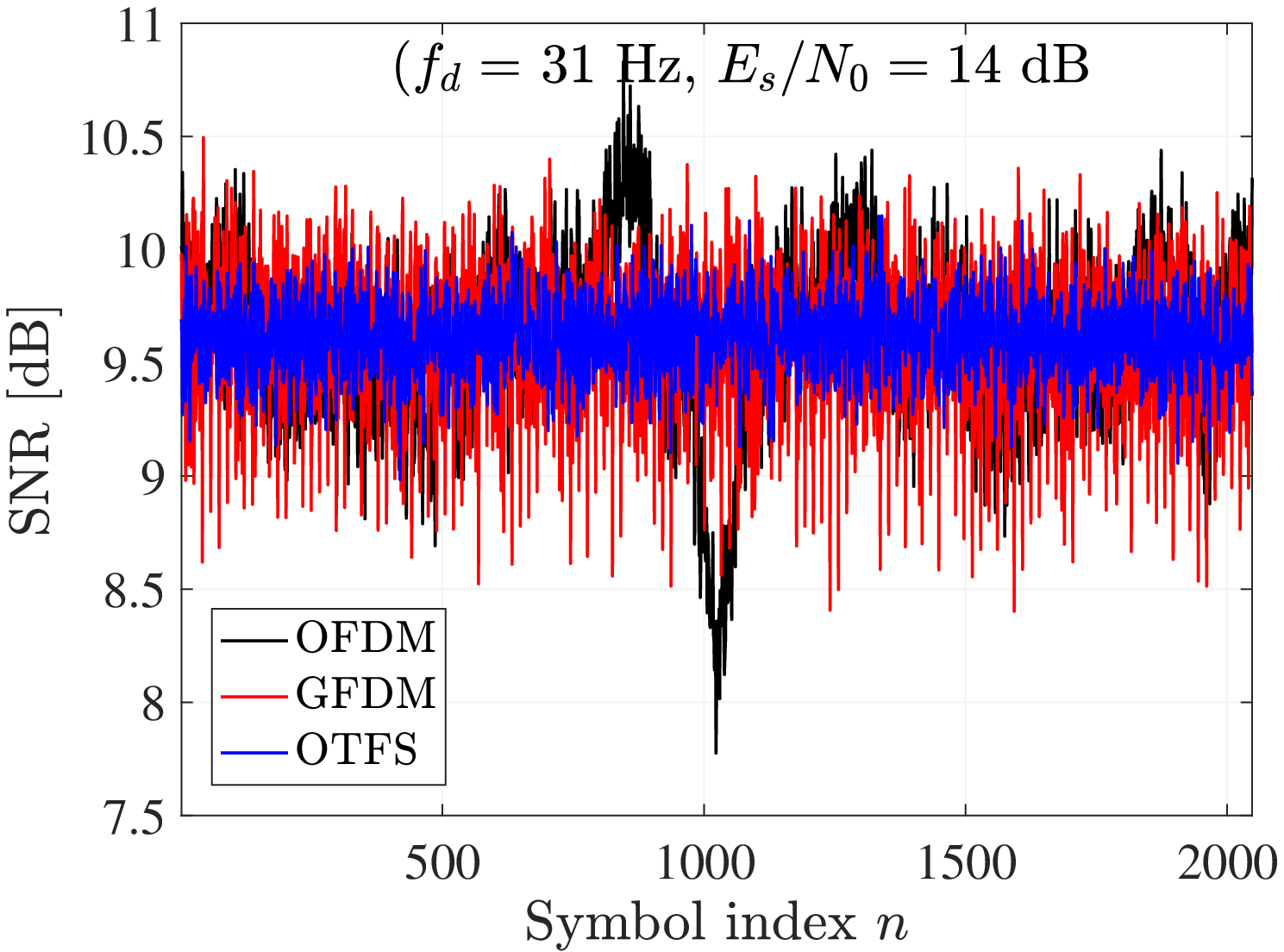}
		\caption{SNR per symbol.}
		\label{fig:per symbol power}
	\end{subfigure}
	\caption{Performance evaluation of \ac{OFDM}, \ac{GFDM} and \ac{OTFS} with long frames.}\label{fig:long_f}
\end{figure*}
\begin{figure*}
	\centering
	\begin{subfigure}[b]{0.31\textwidth}
		\includegraphics[width=\textwidth]{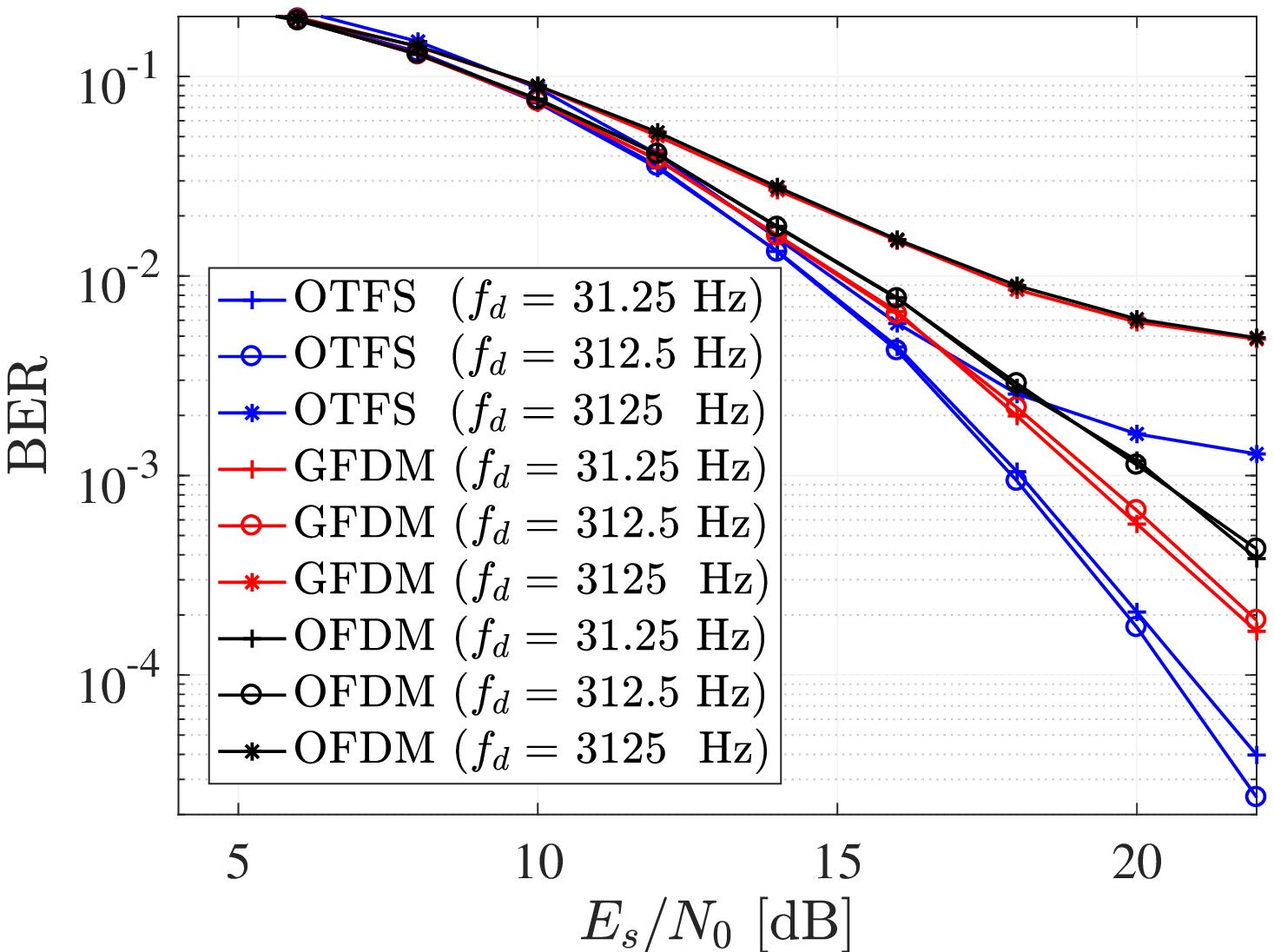}
		\caption{BER vs. SNR.}
		\label{fig:BER_s}
	\end{subfigure}
	~ 
	\begin{subfigure}[b]{0.31\textwidth}
		\includegraphics[width=\textwidth]{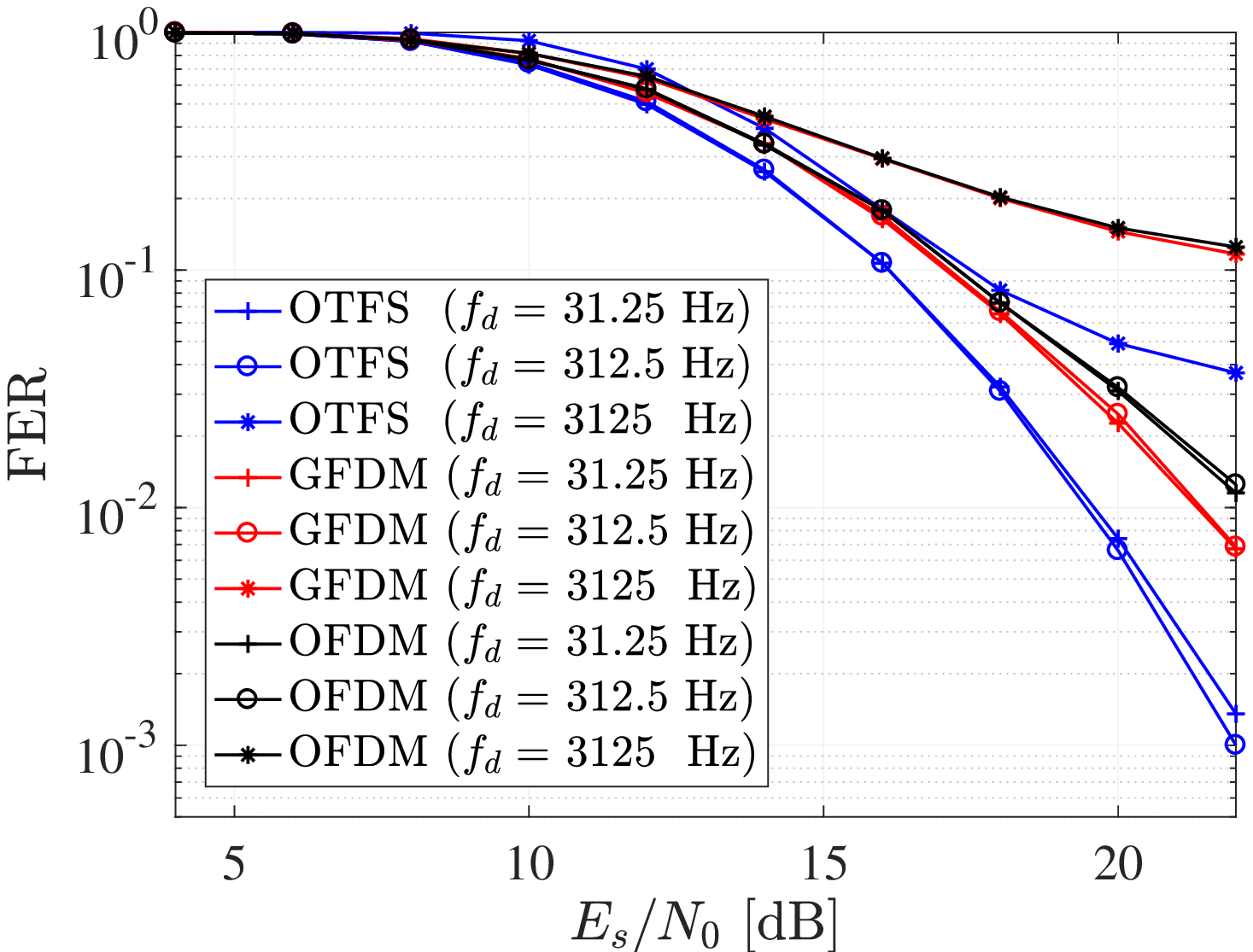}
		\caption{FER vs. SNR.}
		\label{fig:FER_s}
	\end{subfigure}
	~ 
	\begin{subfigure}[b]{0.31\textwidth}
		\includegraphics[width=\textwidth]{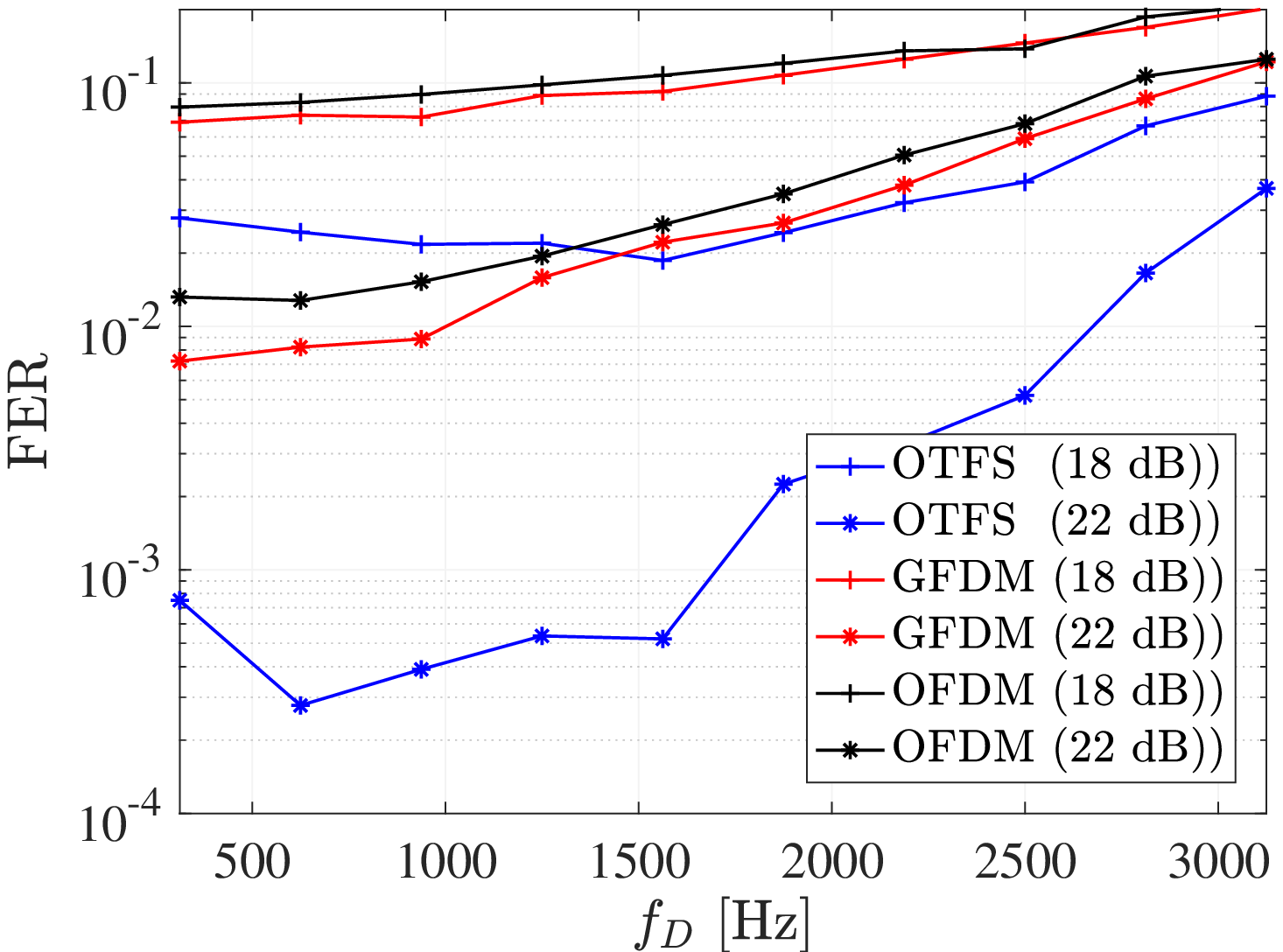}
		\caption{FER vs. Doppler frequency.}
		\label{fig:FER doppler}
	\end{subfigure}
	\caption{Performance evaluation of \ac{OFDM}, \ac{GFDM} and \ac{OTFS} with short frames.}\label{fig:short_f}
\end{figure*}

\section{Numerical simulation}\label{sec:numerical example }
The purpose of this section is first to compare the performance of the conventional \ac{GFDM}\footnote{In this section, \ac{GFDM} refers to the conventional \ac{GFDM} waveform \cite{GFDM}. } and \ac{OTFS} systems  generated from the same parameters. The simulation parameters are listed in Table~\ref{tab:configurations}. The data bits are encoded with \ac{LTE}-Turbo code of code rate $1/2$ and mapped to $16$-{QAM} symbols. The transmit window is generated from a prototype pulse $g$, which is a periodic \ac{RC} with roll-off factor $\alpha = 0$. The samples in both systems are generated from the \ac{GFDM} modulator depicted in Fig.~\ref{fig:Modulator } using the parameters $K=N_o$ and $M=M_o$. The \ac{MMSE} receiver in the case of \ac{OTFS} is performed using \ac{MMSE} pulse shape generated based on the equivalent window $\ma{W}^{(e)}_{\text{tx}}$ \eqref{eq:eq_window}, while the \ac{GFDM}-\ac{MMSE} receiver considers the linear model \cite{zhang2017study}
\begin{equation}
\small
\tilde{\ma{y}} = \diag{\tilde{\ma{h}}^{(e)}}\tilde{\ma{A}}\ma{d}  + \tilde{\ma{\epsilon}}_d +\tilde{\ma{v}}.
\end{equation}
The demodulated symbols are fed to soft-input decoder.
Both modulation techniques are evaluated in the \ac{EVA} \cite{EVA-model} with the consideration of ideal channel estimation per block as in \eqref{eq:ideal CE} in the appendix.
\begin{table}[h]
	\centering
		\small
\caption{Simulation parameters}\label{tab:configurations}
\begin{tabular}{l|l}
   \ac{OTFS}& $N_o= 16,~M_o = 128$ \\ 
	\hline 
 Long-\ac{GFDM}&  $K= 16,~M = 128$ \\ 
\hline
Long-\ac{OFDM}&  $N= 2048$\\ 
\hline 
Short-\ac{GFDM}&  $K= 16,~M = 8$ \\ 
\hline
Short-\ac{OFDM}&  $N= 128$\\ 
	\hline 
	\ac{CP} length $N_{\text{CP}}$& $32$	  \\ 
	\hline 
		Bandwidth $B = F_s$ & $8$ \unit{MHz}	  \\ 
			\hline 
				\ac{OTFS} subcarrier spacing $\Delta f_o$ & $62.5$ \unit{KHz}	  \\ 
			\hline 
Modulation and coding &$16$-QAM, LTE-Turbo $(1/2)$ \\		
 \hline
\end{tabular} 
\end{table}
The \ac{SNR} is defined by the ratio $E_s/N_0$, where $E_s$ is the average symbol power and $N_0$ is the additive white noise power.
\begin{equation}
\small
\text{NMSE} =\frac{\Ex[]{\NORM{\hat{\ma{d}}-\ma{d}}^2}}{\Ex[]{\NORM{\ma{d}}^2}},~\text{SNR}[n] =\frac{\Ex[]{|\hat{{d}_n}-{d}_n|^2}}{\Ex[]{|{d}_n}|^2}.
\end{equation}
Fig.~\ref{fig:long_f} shows a comparison between \ac{OTFS}, conventional \ac{GFDM} and \ac{OFDM} using long block. A frame is represented in a code word corresponding to one \ac{GFDM} block. It is shown that, the Doppler has less influence on \ac{OTFS} compared to the \ac{GFDM} and \ac{OFDM}. As discussed in the appendix, this is because of the accuracy of the equivalent channel estimation, which increases with the decrease of the product  $P{\nu_d}$, where $\nu_d = \frac{f_d}{B}$ and $P$ is the \ac{DFT} size. In \ac{OTFS}, $P = M_o$, while in the long \ac{GFDM} and \ac{OFDM} $P = KM_o$, and thus,  \ac{GFDM} and \ac{OFDM} obviously suffer from higher Doppler interference. This is reflected in the \ac{BER} performance as shown in Fig.~\ref{fig:BER_l}. However, the achieved gain is at the cost of increased \ac{CP} overhead in the case of \ac{OTFS}.  Although \ac{GFDM} and \ac{OFDM} achieve similar performance  in term of \ac{BER} for smaller Doppler shift, \ac{OTFS} significantly outperforms them in term of \ac{FER} as depicted in Fig.~\ref{fig:FER_l}. For instance, for  $f_d = 31.5\unit{Hz}$ the \ac{FER} gain is $3$\unit{dB} at \ac{SNR} $>14$\unit{dB}.  This can be explained, as illustrated in Fig.~\ref{fig:per symbol power},  by the almost equal  \ac{SNR} per-symbols at the output of the \ac{OTFS} demodulator unlike in \ac{GFDM} and \ac{OFDM}, where the \ac{SNR} per symbol experiences significant variations. Additionally, these variations are even higher in \ac{OFDM} such that \ac{GFDM} achieves slightly lower \ac{FER} than \ac{OFDM}. Based on that, the corresponding bit errors in \ac{OFDM} and \ac{GFDM} can be burst, which increases the number of frame errors. But in the case of \ac{OTFS}, it is more likely that  an erroneous frame has higher number of bit errors. On the other hand, it can be observed that, with increased Doppler shift, e.g. $f_d = 312.5\unit{Hz}$, \ac{OTFS} maintains smaller \ac{FER} through the exploitation of time and frequency diversity, while the \ac{GFDM} and \ac{OFDM} links are almost broken. Furthermore, the Doppler interference increases with the increase of the transmit power, which explains the \ac{BER} floor.

However, the design of \ac{OFDM} and \ac{GFDM} can be updated with respect to the channel mobility conditions. Therefore, another configuration considering shorter block length of \ac{GFDM} and \ac{OFDM} is evaluated in Fig.~\ref{fig:short_f}. Moreover, the frame length in \ac{OTFS}, i.e. the code word is shortened accordingly. In this scenario, \ac{GFDM} and \ac{OFDM} attain better performance due to the lower Doppler interference but less robust than \ac{OTFS} against increased Doppler frequencies as illustrated in Fig.~\ref{fig:BER_s}. Moreover, due to the spreading technique in \ac{OTFS}, the short frame is spread over larger time duration. As a result, the \ac{FER} performance gain of \ac{OTFS} is appealing as shown in Fig.~\ref{fig:FER_s}, especially at high Doppler shift. For instance, the gain is higher than $4$ \unit{dB} at $f_d>1500$\unit{Hz} an SNR $ = 18$\unit{dB}, as can bee seen in Fig.~\ref{fig:FER doppler}. On the other hand, considering additional low-latency and throughput constraints, the conventional \ac{GFDM} provides acceptable performance under moderate mobility conditions.
\section{Conclusion}\label{sec:conclusions}
In this paper, we have proposed an extended \ac{GFDM} framework  that provides unified four-step implementation structure in the time and frequency domains. The \ac{OTFS} modulation, which is a powerful transmission technique in \ac{LTV} channels, can be processed with the \ac{GFDM} framework with simple modification. Namely, it is shown that the \ac{OTFS} block results from a permutation of the corresponding conventional \ac{GFDM} block and the received \ac{OTFS} block can be represented as a \ac{GFDM} block in additive noise channel. Thus, the received \ac{OTFS} signal can be fed to the conventional \ac{GFDM} demodulator with a reconfigurable receive pulse shape that depends on the channel estimation. As a result, the \ac{OTFS} equalization and demodulation are combined in a circular filtering that produces data symbols with equal \ac{SNR}. In addition, the \ac{MMSE}-\ac{OTFS} receiver can be implemented with low complexity. The simulation results show that  \ac{OTFS} outperforms the counterparts long-\ac{GFDM} and long-\ac{OFDM} in terms of \ac{BER} and \ac{FER}. Furthermore, \ac{OTFS} significantly outperforms the short blocks of \ac{GFDM} and \ac{OFDM} in high mobility scenarios, due to the exploitation of time and frequency diversity. However, \ac{GFDM} can be a reasonable choice for moderate mobility and low-latency use cases as it outperforms \ac{OFDM} in term of \ac{FER}.

In the future work, the extended \ac{GFDM} framework is used to develop alternative forms of \ac{OTFS} with applications for multiuser scenarios.

\bibliographystyle{IEEEtran}
\bibliography{references}
\appendix
\small
\section{ LTV channel}
Consider a signal $x^{(cp)}[n],~ n= 0\cdots N+N_{\text{cp}}-1$ with $x^{(cp)}[n] = x[<n-N_{\text{cp}}>_N]$ transmitted through \ac{LTV} channel defined by the delay-Doppler response $H(l,\nu)$, where $\nu$ is the normalized Doppler frequency, i.e. $\nu = \frac{f}{F_s}$, the received signal can be expressed as  
\begin{equation}
\small
\begin{split}
r[n] = \int_{\nu} \sum_{l=0}^{L-1}H(l,\nu) x^{(cp)}[n-l]e^{j2\pi n\nu}d\nu,
\end{split}
\end{equation}
with $L\leq N_{\text{cp}}-1$ is the maximum delay spread. Then we extract the samples $y[n]$ as $y[n]=r[N_{\text{cp}} +n],~ n=0\cdots N-1$,
\begin{equation}
\small
\begin{split}
\mbox{then~~} y[n]&= \int_{\nu} \sum_{l=0}^{L-1}H(l,\nu) x[<n-l>_N]e^{j2\pi [N_{\text{cp}} +n]\nu}d\nu\\
&=\frac{1}{N}\int_{\nu} \sum_{k=0}^{N-1}\tilde{H}(k,\nu) \tilde{x}[k]e^{j2\pi\frac{kn}{N}}e^{j2\pi [N_{\text{cp}}+n ]\nu}d\nu.
\end{split}
\end{equation}
Here, $\tilde{H}(k,\nu) = \sum_{l=0}^{L-1}H(l,\nu) e^{-j2\pi\frac{kl}{N}}$. This corresponds to removing the \ac{CP} and using $N$-\ac{DFT} representation for circular convolution, which is attained due to the \ac{CP}. The DFT of the received signal is 
\begin{equation*}
\small
\begin{split}
\tilde{y}[q] &=\sum_{n=0}^{N-1}y[n]e^{-j2\pi\frac{nq}{N}}\\
&=\frac{1}{N}\int_{\nu} \sum_{k=0}^{N-1}\tilde{H}(k,\nu) \tilde{x}[k]e^{j2\pi N_{\text{cp}}\nu}\sum_{n=0}^{N-1}e^{j2\pi\frac{n(k-q+N\nu)  }{N}} d\nu.
\end{split}
\end{equation*}
Assuming that the maximum Doppler spread $f_d$ satisfies  $N|\nu_D| \ll 1$, where $\nu_d = \frac{f_d}{F_s}$,  then,
\begin{equation}
\small
\sum_{n=0}^{N-1}e^{j2\pi\frac{(k-q)n}{N}}e^{j2\pi n\nu} \approx\sum_{n=0}^{N-1}e^{j2\pi n\nu}\delta(q-k), \label{eq: approx}
\end{equation}
\begin{equation}
\small
\begin{split}
\tilde{y}[q] 
&\approx \frac{1}{N}\sum_{n=0}^{N-1}\int_{\nu} \tilde{H}(q,\nu) \tilde{x}[q]e^{j2\pi N_{\text{cp}}\nu}e^{j2\pi n\nu}d\nu\\
&= \frac{1}{N}\sum_{n=0}^{N-1} \tilde{h}(q,N_{\text{cp}}+n) \tilde{x}[q].
\end{split}
\end{equation}
Thus, the equivalent channel vector $\tilde{\ma{h}}^{(e)}$ in the frequency domain is  
\begin{equation}
\small
\begin{split}
\IndexV{\tilde{\ma{h}}^{(e)}}{q} &= \frac{1}{N}\sum_{n=0}^{N-1} \tilde{h}(q,N_{\text{cp}}+n)\\
&= \frac{1}{N}\sum_{n=0}^{N-1} \sum_{l=0}^{L-1}h(l,N_{\text{cp}}+n) e^{-j2\pi\frac{ql}{N}} 
\end{split},
\end{equation}
where $h(l,n)$  is the channel impulse response.
This corresponds to averaging over the time instance. As a result, 
\begin{equation}
\small
\tilde{\ma{y}} = \tilde{\ma{h}}^{(e)}\odot  \tilde{\ma{x}} + {\ma{\epsilon}_d},
\end{equation}
where ${\ma{\epsilon}_d}$ represents the  interference due to the Doppler spectrum and depends on the accuracy of the approximation in  \eqref{eq: approx}. 
The equivalent channel impulse response ${{h}}^{(e)}[n]$ can be estimated by sending a Dirac pulse through the channel as
\begin{equation}
\small
h^{(e)} [n] = \int_{\nu}\sum\limits_{l=0}^{L-1}H(l,\nu)\delta(n-l)e^{j2\pi n\nu} d\nu. \label{eq:ideal CE}
\end{equation}
This model can be used as an ideal channel estimation.

\end{document}